\documentclass[prl,superscriptaddress,twocolumn,
 amsmath,amssymb
]{revtex4-2}

\usepackage{graphicx}
\usepackage{dcolumn}
\usepackage{bm}
\usepackage{hyperref}
\usepackage[mathlines]{lineno}
\usepackage{xcolor}
\usepackage{sidecap}
\usepackage{url}

\usepackage{natbib}

\begin{document}


\title{Selection for size in molecular self-assembly drives the \textit{de novo} evolution of a molecular machine}

\author{Zena Hadjivasiliou}\email{zena.hadjivasiliou@ucl.ac.uk}
\affiliation{Department of Physics and Astronomy, University College London, United Kingdom}
\affiliation{Institute for the Physics of Living Systems, University College London, United Kingdom}
\affiliation{Mathematical and Physical Biology Laboratory, The Francis Crick Institute, United Kingdom}
\affiliation{Department of Biochemistry, University of Geneva, Geneva, Switzerland}
\author{Karsten Kruse}\email{karsten.kruse@unige.ch}
\affiliation{Department of Biochemistry, University of Geneva, Geneva, Switzerland}
\affiliation{NCCR for Chemical Biology, University of Geneva, Geneva, Switzerland}
\affiliation{Department of Theoretical Physics, University of Geneva, Geneva, Switzerland}

%



\date{\today}

\begin{center}
\begin{abstract}
The functioning of machines typically requires a concerted action of their parts. This requirement also holds for molecular motors that drive vital cellular processes and imposes constraints on their conformational changes as well as the rates at which they occur. It remains unclear whether features required for functional molecular machines can emerge simultaneously or require sequential adaptation to different selection pressures during evolution. We address this question by theoretically analyzing the evolution of filament treadmilling. This process refers to the self-assembly of linear polymers that grow and shrink at equal rates at their opposite ends. It constitutes a simple biological molecular machine that is notably involved in bacterial cell division and requires that several conditions are met. In our simulation framework, treadmilling emerges as a consequence of selecting for a target average polymer length. We discuss, why other forms of assembly dynamics, which also reach the imposed target length, do not evolve in our simulations. Our work shows that complex molecular functions can evolve \textit{de novo} under selection for a single physical feature.

\end{abstract}
\end{center}

\maketitle



A plethora of well crafted molecular machines underlie functions that are fundamental to Life, from DNA replication to transport of cellular cargo, and cell migration~\cite{Leman13thereplication,ROSS200841,DaviBeeby2022}. These processes involve molecular motors that produce directional motion, like myosins and kinesins. Physics imposes that the generation of directional motion at a molecular level requires structural polarity and a departure from thermodynamic equilibrium~\cite{Julicher:1997tu}. How molecular self-assemblies evolved the capacity to form structures that perform robust and precise functions in noisy environments remains largely unknown. In particular, we still understand little about the selection forces that drove the evolution of molecular machines in the first place.

The broader question of how complex organs and mechanisms evolve through natural selection has puzzled evolutionary biologists since Darwin~\cite{darwin1859}. It is often assumed that complexity emerges incrementally so that  different features required for an elaborate function are acquired sequentially due to different selection pressures~\cite{Nilsson2009,DaviBeeby2022}. Here we ask whether a complex molecular machine can evolve as a response to selection for a single  attribute without the need to incrementally achieve features that are necessary for the functioning of that machine. 

\begin{figure*}
\includegraphics[width=\textwidth]{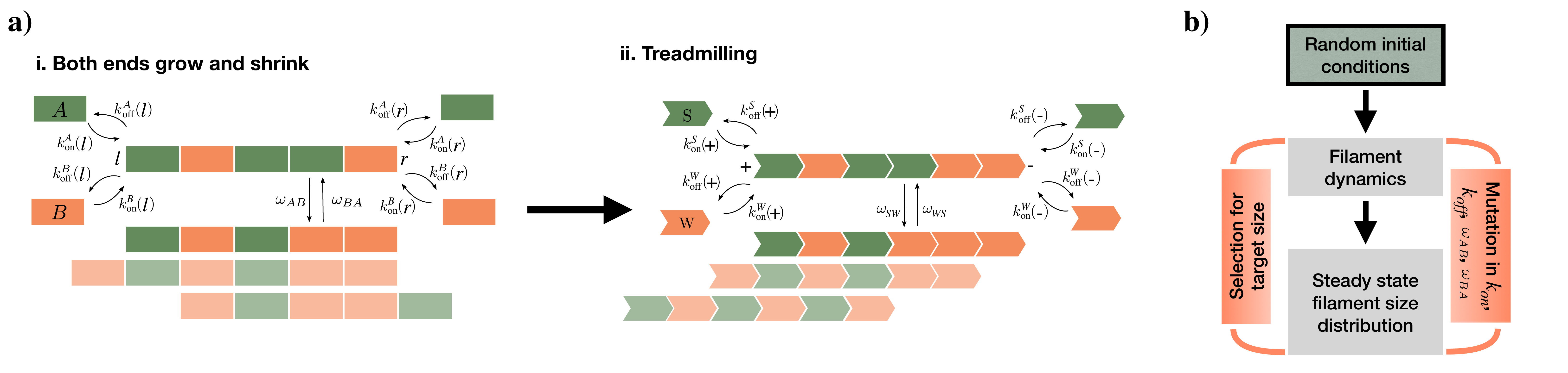}
\caption{\textbf{Schematic diagram of filament dynamics and evolution algorithm.} a) i. A two-state model for active filaments without assuming polarity. Subunits are in one of two states denoted by $A$ (green blocks) and $B$ (orange blocks), respectively. Subunits are added and removed from the two filament ends, $l$ and $r$, at given rates. Transitions between the two subunit states within a filament occur at rates $\omega_{AB}$ and $\omega_{BA}$. ii. Treadmilling is characterised by rates that lead to net growth at one (plus) and net shrinkage at the other (minus) end so that the filament effectively moves in one direction. Here, $S$ and $W$ denote strongly and weakly bound subunit types. b) Diagram of the evolution algorithm employed in this study.}
\label{fig:diagram}
\end{figure*}

Compared to the functioning of motors like kinesins or myosins, filament treadmilling provides a simpler way to generate directional motion~\cite{Bisson2017,Yang2017,McCausland:2021}. Filaments are linear aggregates of identical units such as actin and tubulin in eukaryotic cells as well as their respective bacterial analogs MreB or MamK and FtsZ or TubZ~\cite{Lowe:2009eu}. When treadmilling, units are added to one end and removed from the opposite one at equal rates, Fig.~\ref{fig:diagram}a. This process has been observed for actin filaments~\cite{Wegner:1976ks} and microtubules~\cite{Walker:1988a,rodionov:1997} and plays a particularly important role in bacteria that lack kinesins and myosins or analogs thereof. For example, treadmilling of TubZ is required for plasmid stability in \textit{Bacillus thuringiensis}~\cite{larsen:2007}, treadmilling of FtsZ for division of \textit{Escherichia coli}~\cite{McCausland:2021} and \textit{Bacillus subtilis}~\cite{whitley:2021}, and treadmilling of MamK for the segregation of magnetosomes in \textit{Magnetospirillum gryphiswaldense}~\cite{Toro-Nahuelpan:2016}. On the molecular level, treadmilling is driven by having a nucleotide tri-phosphate (NTP) bound to the assembling units and subsequent hydrolysis into the corresponding nucleotide di-phosphate (NDP). This reaction concurs with a reduction of its affinity for the aggregate~\cite{Kabsch1990,Hyman1995,Ruiz2022}. When linked to other structures treadmilling can generate mechanical stress. 

In this work, we consider treadmilling to explore possible routes to the evolution of a molecular machine. We specifically ask whether selection for filaments of fixed length could drive the \textit{de novo} evolution of treadmilling filaments, that is, without the requirement that any of the features needed for treadmilling are initially present. To answer this question, we study theoretically the dynamics of linear aggregates of NTP-binding units under mutation and selection. 

We describe a filament as a dynamic linear lattice with two ends (Fig. \ref{fig:diagram}a). Each lattice site represents a single subunit that can be in two  states, denoted $A$ or $B$, corresponding to the different nucleotides that can be bound to actin, tubulin, and their bacterial analogs. Subunits within the lattice can switch between the two states with rate $\omega_{AB}$ from  $A$ to $B$ and with rate $\omega_{BA}$ for the opposite transition. Single subunits can be added to or removed from either end of the lattice. We denote the two ends of the lattice by $l$ and $r$ and define the various rates of subunit addition and removal at the two ends by $k_\mathrm{on}^{A}(l)$, $k_\mathrm{on}^{A}(r)$, $k_\mathrm{off}^{A}(l)$, and $k_\mathrm{off}^{A}(r)$ for subunits in state $A$ and similarly for subunits in state $B$, Fig.~\ref{fig:diagram}a. We assume that the unbound subunits form a reservoir, such that the on-rates do not depend on the lattice length. We also assume that these rates only depend on the state of the subunit that is added to or removed form the lattice. Consequently, the equilibrium constants are equal at the two ends, $K_A\equiv k_\mathrm{on}^{A}(l)/k_\mathrm{off}^{A}(l) = k_\mathrm{on}^{A}(r)/k_\mathrm{off}^{A}(r)$ and similarly for subunits in state $B$~\cite{Erlenkamper:2013cy}. Here, we allow the kinetic parameters that underlie self-assembly to mutate and impose selection for a particular filament length, Fig.~\ref{fig:diagram}b. 

In this description, treadmilling can in principle occur for a broad range of parameter values but requires the rates introduced above to fulfill certain conditions~\cite{Erlenkamper:2013cy}. First, the two subunit states must have different affinities. For treadmilling filaments we refer to the two states as $S$ for strongly and $W$ for weakly bound. Second, net growth at one end, referred to as `plus', and net shrinkage at the `minus' end occurs if $S$ subunits  are more likely to be present at the plus end and \textit{vice versa} for $W$ subunits. Finally,  for a filament of finite but nonzero length, subunits in the $S$ state must be added to the filament more rapidly than they switch to the $W$ state. These conditions lead to a gradient in the probability of finding a subunit in the $S$ state along the lattice length, implying a length-dependent depolymerization rate at the minus end~\cite{Erlenkamper:2013cy}. It follows that filament treadmilling requires the coordination between several independent rates, as well as different features such as polarity and different transition rates between two subunit states.

Importantly, it is not obvious whether the features described above can all emerge together in evolution under the same selection pressures or must evolve sequentially. In addition, a target average filament length is also attainable, when the filament is not treadmilling~\cite{Mohapatra2016}. How  selection for a target average filament length will affect the evolution of the self-assembly dynamics of our system is thus unclear. 

We analyzed the above description through stochastic simulations of the discrete lattice dynamics following the Gillespie scheme, where we introduced a maximum lattice length of 2000 sites. When this number was reached, no further binding was allowed to reduce simulation time. We ran each simulation for $10^6$~s of simulated time to reach steady state. The steady state is expected to be reached within that timescale since the minimum value for all rates was set to $10^{-5}$~s$^{-1}$. Subsequently, we continued the simulations and sampled the lattice length every 20 simulation steps until $10^6$ samples were obtained.

Between runs, we mutated the different rates. The mutation process is intended to capture structural changes of the subunits in states $A$ and $B$ that are reflected by changes of the various rates. To implement mutation, we first changed the ratio $k_\mathrm{on}^{A}(l)/k_\mathrm{off}^{A}(l)$ by a factor $10^f$, where $f$ is sampled from a uniform distribution on the interval $[-\mu, \mu]$. Here $\mu=2$ determined the mutation magnitude. Afterwards we randomly chose one of the two parameters with equal probability to remain constant, whereas the other was modified to reach the change in their ratio determined by the mutation magnitude. Finally, we set the ratio of the on- to the off-rate at the $r$ end to that at the $l$ end by randomly choosing one of the rates at the $r$ end with equal probability and modifying it accordingly. The same procedure was then applied to the rates for $B$-subunits. Finally, we independently mutated the values of $\omega_{AB}$ and $\omega_{BA}$ again by multiplication with a factor 10$^f$ where $f$, is sampled from a uniform distribution on the interval $[-\mu, \mu]$. Each rate is allowed to vary in the interval $[10^{-5}, 1]$~s$^{-1}$.

We implemented selection by accepting mutations that confer a change in the steady state filament length such that $ \mathcal{E} (\mathbf{k}') < \mathcal{E}(\mathbf{k})$. Here, $\mathcal{E}$ is some energy function and $\mathbf{k}$ and $\mathbf{k'}$ are vectors that respectively hold all rates before and after mutation.  We performed a series of rounds of mutation and selection until $\mathcal{E}(\mathbf{k}) < \delta$, Fig.~\ref{fig:diagram}b. For the purposes of the analysis below, we chose $\mathcal{E(L)} = (L-L^*)^2 $ and $\delta= (L^*/20)^2$ where $L$ is the average length of the current parameter set and $L^*=200$~subunits is the target length.

Our evolution algorithm mirrors an adaptive dynamics approach where we assume that the accumulation of small mutations leads to continuous changes in the parameters that specify self-assembly. This approach assumes that  evolution occurs in a large, asexual population and is fast so that new mutations vanish or invade to replace the resident quickly~\cite{AdaptiveD2013}. Positive values in the difference $ \mathcal{E} (\mathbf{k}) - \mathcal{E}(\mathbf{k}')$ reflect a positive selection gradient and lead to the invasion of the mutant. 

\begin{figure}
\centering
\includegraphics[width=8.6cm]{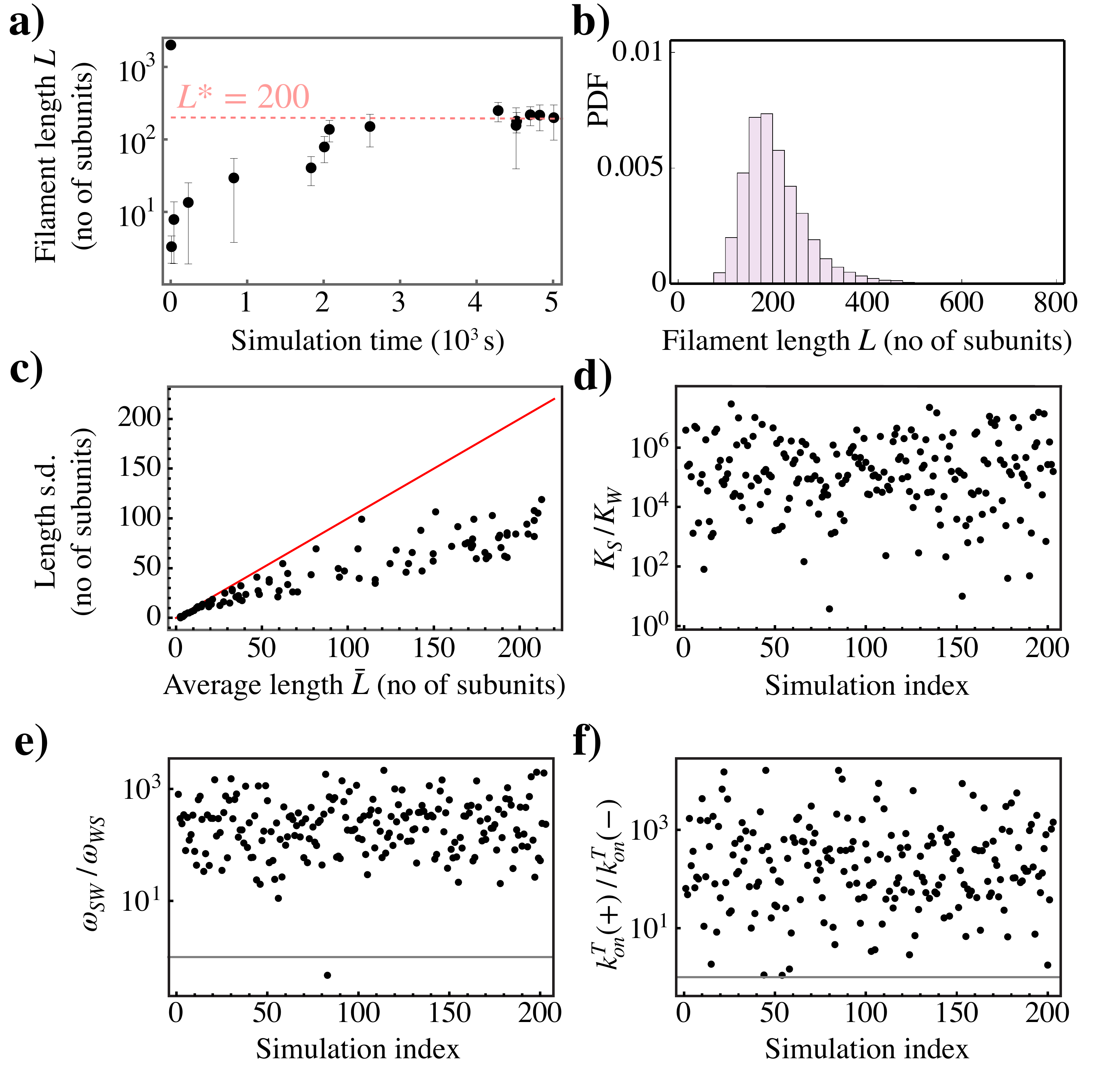}
\caption{\textbf{Filament converges to target length and evolved filaments are polar.} a) Example of filament mean length evolution. Bars indicate standard deviation. 
b) Filament length distribution at mutation-selection balance for a single evolved instance. c) Standard deviation \textit{versus} mean length. Data is pooled for all evolution steps across all simulations. d) Ratio of the binding affinity of $S$-subunits over $W$-subunits. e) Rate at which $S$-subunits transform into $W$-subunits over the rate of the inverse transformation. f) Ratio of the on rates of $S$-subunits at the plus over minus end. }
\label{fig:polarity}
\end{figure}

We start our evolution algorithm from random initial values for the different rates. Initial conditions typically resulted in the lattice reaching maximum length. Through mutation and selection, the average lattice size converged to the target value, Fig.~\ref{fig:polarity}a. The process of convergence to the target mean length was typically biphasic: first the average length decreased to a value of just a few subunits. At this point of the evolution process, the length distribution was exponential. When we continued the evolution process, the average length gradually increased. The corresponding length distribution remained unimodal, Fig.~\ref{fig:polarity}b, but with a reduced standard deviation compared to the mean length, Fig.~\ref{fig:polarity}c. 

In all our simulations ($N=204$), we found that a difference between the equilibrium constants $K_S$ and $K_W$, that is, between the binding affinities of subunits in the two states evolved, Fig.~\ref{fig:polarity}d. 
In the vast majority of our simulations, the difference was more than 2 orders of magnitude. With one exception, the conversion rate from the $S$ to the $W$ state was larger by at least an order of magnitude relative to the opposite process. In addition, all evolved states exhibited polarity, such that the maximal on rate at the plus end substantially exceeded that at the minus end, Fig.~\ref{fig:polarity}f. Together, these results show that selection for an average filament length can lead to the evolution of subunits of different affinities, polar lattices and different transition rates between the two subunit states. 

\begin{figure}[t]
\centering
\includegraphics[width=8.6cm]{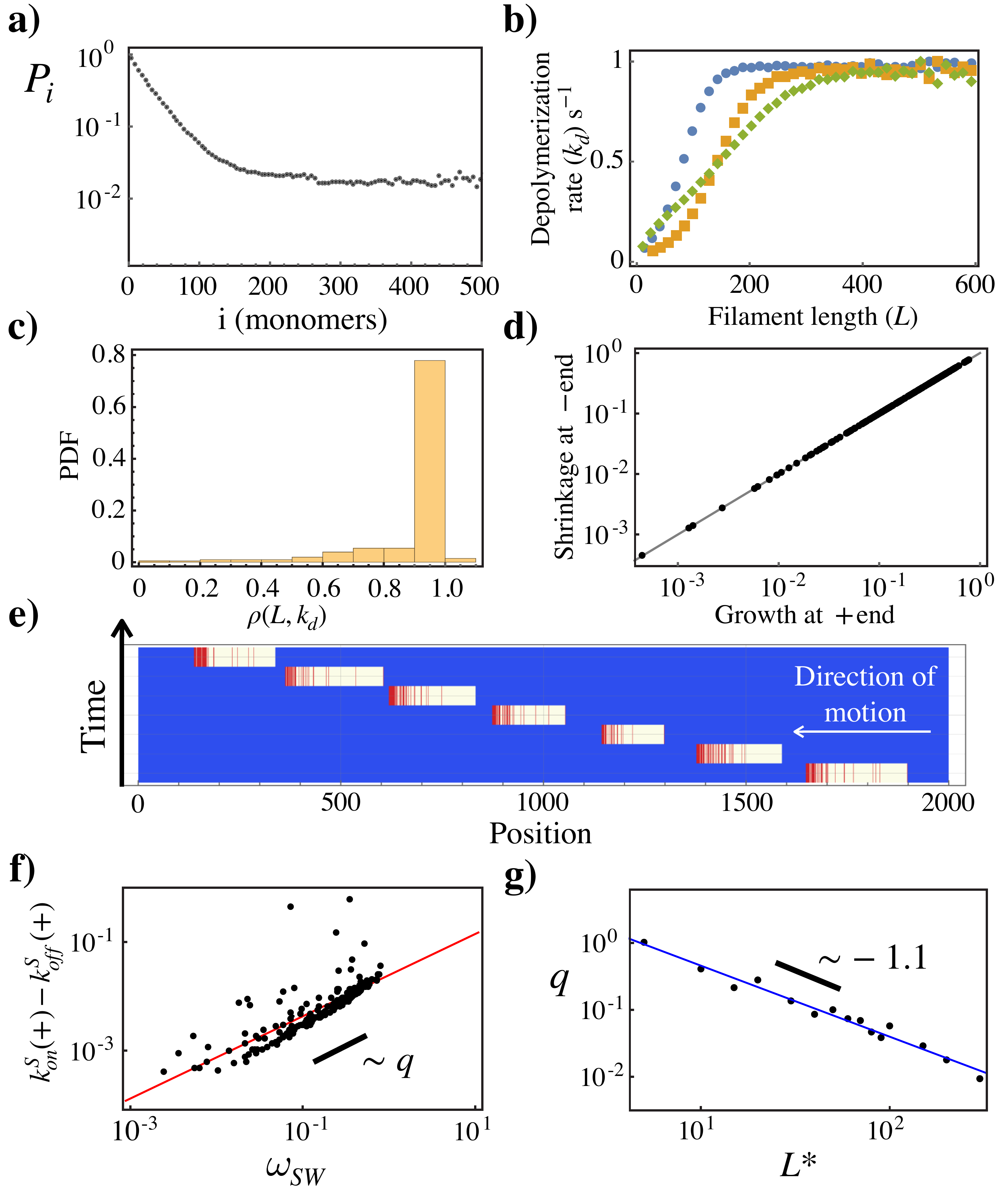}
\caption{\textbf{All evolved states exhibit treadmilling dynamics.} a) Probability $P_i$ for site $i$ to be in the S-state in filaments of length $L>i$ (single evolved instance shown at steady state).  b) Depolymerization rate as a function of filament length (three evolved instances shown at steady state; depolymerization rate is normalized to the maximum).  c) Histogram of the Spearman Correlation coefficient between filament length, $L$, and depolymerisation rate, $k_d$, at the mutation-selection balance for all 204 cases explored in simulations. A value of $\rho$ close to 1 indicates a monotonic relationship between $L$ and $k_d$. d) Net shrinkage rate at the minus end versus net growth rate at plus end at the mutation-selection balance for all 204 cases explored in simulations. Solid line for $y=x$ (correlation coefficient $R^2 = 0.99$). e) Example of treadmilling dynamics. $S$ and $W$ subunits are shown in red and white, respectively, background in blue. Plus end points to the left. f) Difference between the on- and off-rates of $S$-subunits at the plus. 
g) The value of $q$ as a function of the target length.}
\label{fig:TM}
\end{figure}

These properties match the conditions required for treadmilling. To explore whether the lattices exhibit treadmilling dynamics for the evolved parameters, we first measured in our simulations the position-dependent probability for a site to be in the $S$ state. This probability decreased monotonically from the plus end until about twice the target lattice length, Fig.~\ref{fig:TM}a. Beyond this length, the statistics was too weak to obtain a good estimate. In steady state, the gradient in the $S$-state probability led to an effective detachment rate at the minus end that increased monotonicaly with the lattice length, Fig.~\ref{fig:TM}b,c. 

We next computed the effective growth and shrinkage rates of the evolved filaments at steady state. The rate at which lattices grew at the plus end
was equal to the rate at which they shrank at the minus end, confirming that treadmilling dynamics occurs in all evolved states, Fig.~\ref{fig:TM}d. An instance of an evolved filament over time is shown in Fig. \ref{fig:TM}e, illustrating that evolved filaments exhibit  treadmilling and directed motion. Finally, the difference between the on and off rates for  $S$ subunits at the plus end in the evolved filaments increased as a power law with the evolved switching rate $\omega_{SW}$ so that $k_{on}^S(+) - k_{off}^S(+) \sim \omega_{SW}^q$,  Fig.~\ref{fig:TM}f. In addition, varying the target length in the evolution algorithm revealed that $q$ decreased with $L^*$, Fig.~\ref{fig:TM}g. Therefore,  evolution in our algorithm regulates the ratio between $k_{on}^S(+) - k_{off}^S(+)$ and $\omega_{SW}$ to ensure longer residence times prior to detachment for larger filaments. 

As an alternative to treadmilling, the average target length could also be reached for an exponential length distribution. This distribution occurs when growth and shrinkage at the lattice ends are on average independent of the lattice state~\cite{Mohapatra2016,Erlenkamper:2013cy}. That is, if $k_{off}^A(l)p_A(l)+k_{off}^B(l)p_B(l)\equiv k_{off}(l)=const$, where $p_A(l)$ and $p_B(l)$ are the respective probabilities of having an $A$ or a $B$ site at the $l$ end of the lattice, and analogously for the $r$ end. For state-independent effective off rates, the length distribution is given by $P(L)=(1-\alpha)\alpha^L$ for $L=0,1,2,\ldots$, where $\alpha=(k_{on}(l)+k_{on}(r))/(k_{off}(l)+k_{off}(r))$. The average length is then $\alpha/(1-\alpha)$, if $\alpha<1$, such that all possible target average lengths can be reached in this way. However,  this configuration  never  evolved  in our analysis. 

\begin{figure}[t]
\centering
\includegraphics[width=8.6cm]{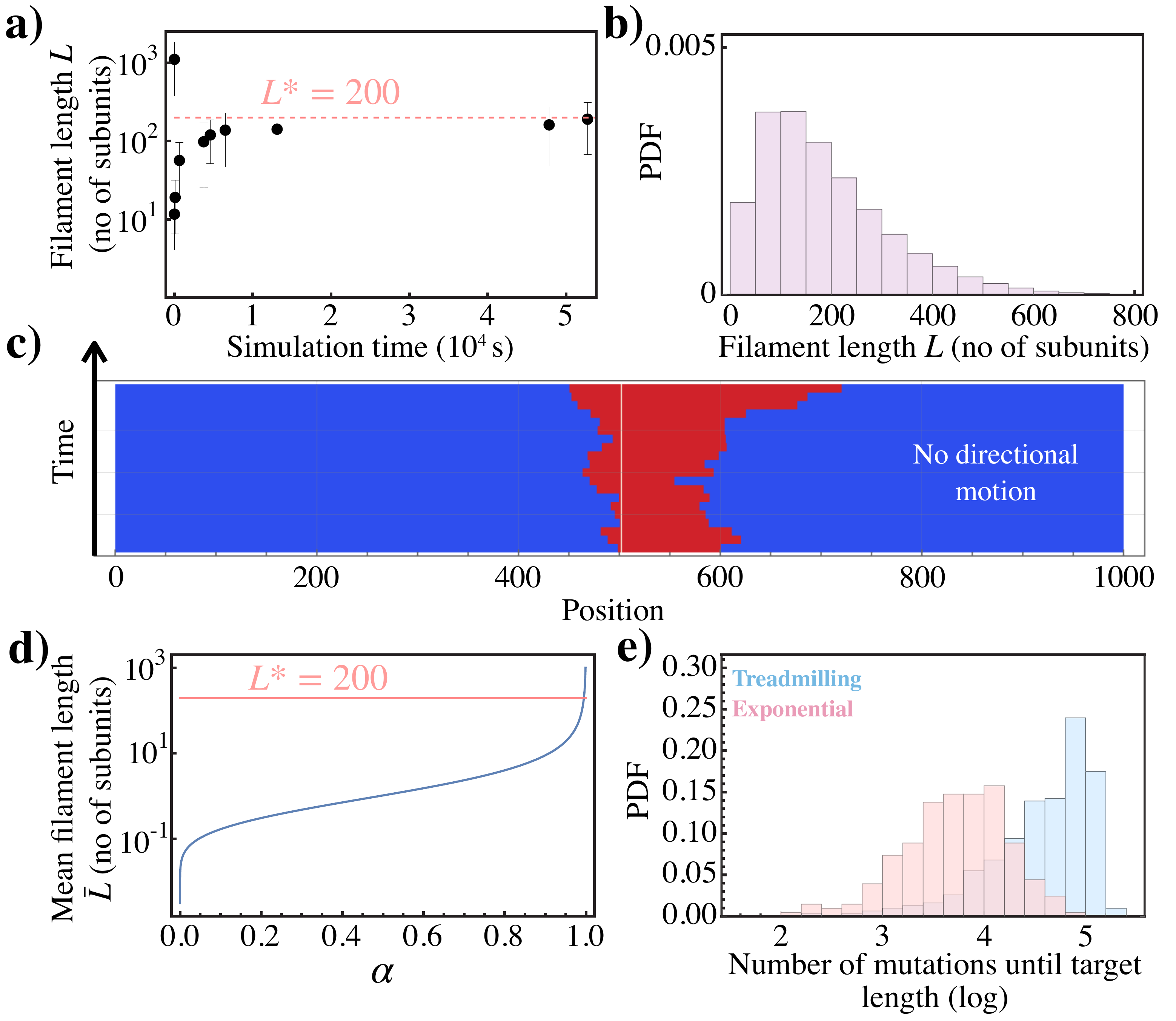}
\caption{\textbf{Evolved exponential distributions are sensitive to binding rates.} a) Example of filament mean length evolution with a single subunit. Bars indicate standard deviation.  b) Filament length distribution at evolutionary equilibrium. c) Typical example of evolved filament dynamics in space. d) Mean filament length as a function of $\alpha$. e) PDFs of the number of mutations required to reach the target length in full (pink) and simplified (blue) simulations. }
\label{fig:exponential}
\end{figure}

We repeated our analysis in a simplified system with only one subunit, $A$. We introduced mutations in the binding and unbinding rates at the two ends and selected for the target length as before. Under these conditions, the filament average length consistently converged to the target length and an exponential length distribution emerged, Fig.~\ref{fig:exponential}a,b.  The evolved filaments grew and shrank independently at both ends and did not exhibit directed motion, Fig.~\ref{fig:exponential}c.

From these simplified dynamics we can infer why exponential distributions did not evolve in our full simulations. Exponential distributions can yield  filament lengths with $L^*\gtrsim10$ only when $\alpha$ is close to 1, i.e. when the subunit addition and removal rates are closely matched, Fig.~\ref{fig:exponential}d. Furthermore, the mean length of an exponential distribution diverges as $\alpha$ approaches 1, Fig.~\ref{fig:exponential}d. Therefore, the mean filament length is extremely sensitive to mutations that change the value of $\alpha$ suggesting that exponential distributions may be less evolvable compared to the treadmilling state. Accordingly, the total number of mutations needed to reach the target filament length through treadmilling in the full simulations was typically more than an order of magnitude smaller than in the simplified simulations, Fig.~\ref{fig:exponential}e. This is particularly striking as only 3 rates are subject to mutation and selection in the latter, whereas there are 8 such parameters in the full simulation.

To  explore the sensitivity of the filament mean length on parameters in treadmilling versus exponential distribution we introduced small perturbations in the binding rates in evolved parameter sets that yield treadmilling dynamics and exponential distributions. As expected, a substantially larger divergence away from the target length for the same shifts in the binding rate is seen in the exponential distribution compared to that obtained under treadmilling, Fig.S1.


In conclusion, considering the specific case of filament treadmilling, we have explored the possibility to evolve molecular building blocks into machines that can perform work by selecting for a single physical feature. Physical constraints are recognised as central players in evolution~\cite{thompson_1992,cockell2018}, but how such constraints can lead to new forms and functionalities is poorly understood. Physical aspects have been included in evolutionary algorithms to organize contractile and passive blocks into clusters that can operate functions~\cite{KriegmanEtAl2020}. This approach has also been exploited to elucidate design rules underlying nanoparticle uptake by cells~\cite{FortserEtAl2020}. Our findings extend these works as they suggest that physical constraints governing molecular self-assembly can in principle lead to the \textit{de novo} evolution of complex functions and molecular machines. This work does not directly address the evolutionary path through which treadmilling may have emerged in existing polymers, but in the future it will be interesting to explore the implications of our findings for the evolution of early Life. 

\acknowledgments{\textit{Acknowledgments.} ZH was supported by the Francis Crick Institute, which receives its core funding from Cancer Research UK; the UK Medical Research Council and Wellcome Trust. We thank Charlotte Aumeier, Jean-Pierre Eckmann, and Michel Milinkovic for comments on this manuscript.}



\bibliography{bibliography.bib}

\end{document}